\documentclass[aps,showpacs,pra,twocolumn]{revtex4}
\usepackage{mathrsfs} 
\usepackage{amssymb}

\usepackage{graphicx}
\usepackage{dcolumn}
\usepackage{bm}

\begin{document}

\title{Violation of Leggett-Garg inequalities in single quantum dot}
\author{Yong-Nan Sun, Yang Zou, Rong-Chun Ge, Jian-Shun Tang, Chuan-Feng Li $\footnote{email: cfli@ustc.edu.cn}$, and
Guang-Can Guo} \affiliation{Key Laboratory of Quantum Information,
University of Science and Technology of China, CAS, Hefei, 230026,
People's Republic of China}
\date{\today }


\begin{abstract}
We investigate the violation of Leggett-Garg (LG) inequalities in
quantum dots with the stationarity assumption. By comparing two
types of LG inequalities, we find a better one which is easier to be
tested in experiment. In addition, we show that the fine-structure
splitting, background noise and temperature of quantum dots all
influence the violation of LG inequalities.
\end{abstract}


\maketitle 
\bibliographystyle{prsty}
The transition from the strange quantum description to our familiar
classical description is one of the fundamental questions in
understanding the world. This issue was first investigated by
Leggett and Garg in 1985~\cite{1} and led to the formulation of the
so called temporal Bell inequalities. Instead of the correlations
between states of two spatially separated system, they cared about
the correlations of the state of a single system at different time.
What's more, in order to get the inequalities, Leggett and Garg made
two general assumptions~\cite{2}: (1) any two-level macroscopic
system will be at any time in one of the two accessible states
(Macroscopic realism). (2) the actual state of the system can be
determined with arbitrarily small perturbation of its subsequent
dynamics (Non-invasive measurability). Leggett-Garg (LG)
inequalities provide a criterion to characterize the boundary
between the quantum realm and classical one and the possibility of
identifying the macroscopic quantum coherence.

Due to the assumption of noninvasive measurement, which describes
the ability to determine the state of the system without any
influence on its subsequent dynamics, it is difficult to test the LG
inequalities experimentally. There have been many proposals to test
such kind of inequalities by employing superconducting quantum
interference devices~\cite{1,3,4}, but the extreme difficulty of
experiments with truly macroscopic systems, as well as the request
of noninvasive measurement, has made it impossible to draw any clear
conclusion. The LG inequalities have been tested in optical system
with CNOT gate~\cite{5}. A more recent approach takes advantage of
weak measurement~\cite{6,7,8}, in which the dynamics of the system
is slightly disturbed. In this work, we follow a different approach.
Testable LG inequalities is derived by replacing the noninvasive
measurement assumption with the stationarity assumption~\cite{9,10}.

Considering an observable $Q(t)$ of a two level-system such as the
polarization of the photon, in this case, we can define $Q(t)=1$
when the state of the photon is $|H\rangle$, and $Q(t)=-1$ when the
state of the photon is $|V\rangle$. The autocorrelation function of
this observable is defined as $K(t_{1},t_{2})=\langle
Q(t_{1})Q(t_{2}) \rangle$. For three different times $t_{1}$ ,
$t_{2}$ and $t_{3}$ $(t_{1}<t_{2}<t_{3})$ , we get the two LG
inequalities under the realistic description~\cite{1}
\begin{equation}
K(t_{1},t_{2})+K(t_{2},t_{3})+K(t_{1},t_{3})\geq -1 ,
\end{equation}
\begin{equation}
K(t_{1},t_{3})-K(t_{1},t_{2})-K(t_{2},t_{3})\geq -1 .
\end{equation}
Once the assumption of stationarity is introduced, in the case of
$t_{2}-t_{1}=t_{3}-t_{2}=t$ , the inequalities become~\cite{9}
\begin{equation}
K_{+}=K(2t)+2K(t)\geq-1 ,
\end{equation}
\begin{equation}
K_{-}=K(2t)-2K(t)\geq-1 .
\end{equation}

These inequalities set the boundary of the temporal correlations and
are amenable to an experimental testing by a two-shot experiment.

In this paper, we compare the two types of LG inequalities and
discuss the violation of LG inequalities with the influence of the
fine-structure splitting, background noise and temperature in a
quantum dot system.

Semiconductor quantum dots, often referred to as ``artificial
atoms'', have well defined discrete energy levels~\cite{11} due to
their three-dimensional confinement of electrons. The atomlike
properties of the single semiconductor quantum dot, together with
its ease of integration into more complicated device structures,
have made it an attractive and widely-studied system for
applications in quantum information~\cite{12}. One of their
potentially useful properties is the emission of polarization
entangled photon pairs by the radiative decay of the biexciton
state~\cite{13,14,15,16}.

\begin{figure}
\includegraphics[width=3in]{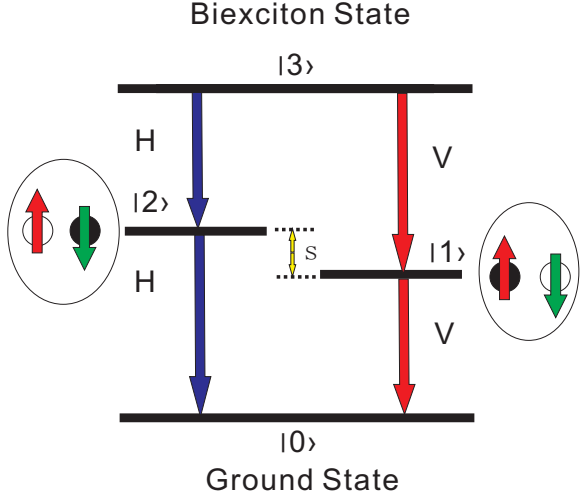}
\caption{(Color online). Energy-level schematic of the biexciton
cascade process. The ground state $|0\rangle$, the two
linear-polarized exciton state $|2\rangle$ and $|1\rangle$, the
biexciton state $|3\rangle$, the fine-structure splitting S.}
\end{figure}

The energy level of a quantum dot is shown in Figure 1, we care
about the two exciton states $|2\rangle$ and $|1\rangle$ in a
quantum dot. We define $Q(t)=1$ when the quantum dot state is
$\frac{~1}{\sqrt{2}}(|2\rangle+|1\rangle)$ and $Q(t)=-1$ when the
quantum dot state is $\frac{~1}{\sqrt{2}}(|2\rangle-|1\rangle)$. The
quantum dot is initially excited to the biexciton state by a
short-pulsed laser and then evolves in a thermal bath. After
emitting a biexciton photon, the proposed single quantum dot system
will be in an entangled photon-exciton state. For an ideal quantum
dot with degenerate intermediate exciton states $|2\rangle$ and
$|1\rangle$, it is a maximum entangled state $ |\psi
\rangle=\frac{~1}{\sqrt{2}}[|H \rangle |2 \rangle+|V \rangle |1
\rangle ] $. Then if we find the biexciton photon in state
$\frac{~1}{\sqrt{2}}(|H\rangle+|V\rangle)$, the state of the quantum
dot system will be $\frac{~1}{\sqrt{2}}(|2\rangle+|1\rangle)$. With
the help of measurement, we can prepare the initial state of the
quantum dot system. After a period of time of evolution, the quantum
dot system will emit the second exciton photon. Then, if the state
of the second photon is detected to be
$\frac{~1}{\sqrt{2}}(|H\rangle-|V\rangle)$, it indicates the quantum
dot system ends up in $\frac{~1}{\sqrt{2}}(|2\rangle - |1\rangle)$.
In this case, because of the relationship between photon and the
quantum dot system, we can detect the photon state to evaluate the
LG inequalities. With these photon states, we can evaluate the LG
inequalities.

Taking the acoustic phonon-assisted transition process into account,
the density matrix of this four-level system can be described with
the master equation~\cite{17}
\begin{eqnarray}
\dot{\hat{\rho}}=-i[\hat{H_{0}},\hat{\rho}]+L(\hat{\rho}),
\end{eqnarray}
where $\hat{H_{0}}=\sum_{i=0}^{3} \omega_{i}|i \rangle \langle i|$
and $L(\hat{\rho})$ is the Lindblad term denotes the dissipation
term~\cite{18}. When projected onto the subspace spanned by the four
basis
$\{|H_{1}H_{2}\rangle,|H_{1}V_{2}\rangle,|V_{1}H_{2}\rangle,|V_{1}V_{2}\rangle\}$,
we can get the corresponding two-photon polarization density matrix
$\hat{\rho}_{_{pol}}$. With the presence of the background noise and
the overlap of the photon pairs frequency distributions, the total
density matrix of the photon pairs is divided into three
parts~\cite{18}
\begin{eqnarray}
\hat{\rho}_{_{tot}}=\frac{1}{1+g}[\eta\hat{\rho}_{_{pol}}+(1-\eta)\hat{\rho}_{_{noc}}+g\hat{\rho}_{_{noise}}],
\end{eqnarray}
where $\hat{\rho}_{_{noc}}$ arises from the fine-structure splitting
induced distinguishability between the two decay paths, representing
the nonoverlap of the photon frequency distributions. The third term
$\hat{\rho}_{_{noise}}$, which describes the background noise, is
set as an identity matrix. At last, the total density matrix of the
photon pairs can be expressed as follows
\begin{displaymath}
\mathbf{\hat{\rho}_{_{tot}}}= \left(
\begin{array}{cccc}
 \rho_{11} & 0 & 0 & \rho_{14} \\
 0 & \rho_{22} & 0 & 0 \\
 0 & 0 & \rho_{33} & 0 \\
 \rho_{41} & 0 & 0 & \rho_{44}
\end{array}
\right)
\end{displaymath}

In our model, we can define the observable $Q(t)=1$ when the state
of the photon is
$|+\rangle=\frac{~1}{\sqrt{2}}(|H\rangle+|V\rangle)$, and $Q(t)=-1$
when the state of the photon is
$|-\rangle=\frac{~1}{\sqrt{2}}(|H\rangle-|V\rangle)$. At the time
$t=0$, we detect the first photon under the bases $|+\rangle$ and
postselect the result of $|+\rangle$ to prepare the initial state of
the quantum dot system. After a period of time $t$, we detect the
second photon under the bases $|+\rangle$ and $|-\rangle$. We define
the joint probability $P_{++}$ when the state of the second photon
is $|+\rangle$ and the probability $P_{+-}$ when the state of the
second photon is $|-\rangle$. We detect the photon pairs with delays
$\tau$ in the range $t\leq \tau\leq (t+\omega)$, by employing a
single timing gate~\cite{19}, where $t$ is the start time of the
gate. At last, we can evaluate the autocorrelation by the expression
\begin{eqnarray}
K(t)=P_{++}(t)-P_{+-}(t) .
\end{eqnarray}
The autocorrelation will be measured two times by two independent
experiments which begin with a primary system in an identical
initial state and evolves under identical conditions. In the first
experiment, the measurement of the autocorrelation is made at the
time $t$ and the second measurement is made at the time $2t$ by
another experiment and the joint probability can be calculated
through the total density matrix of the photon pairs, then the final
autocorrelation can be expressed as follows
\begin{eqnarray}
K(t)=\frac{\rho_{14}(t)+\rho_{41}(t)}{\rho_{11}(t)+\rho_{22}(t)+\rho_{33}(t)+\rho_{44}(t)}
,
\end{eqnarray}
Therefore, we can evaluate the LG inequalities through the
expression $K_{+}=K(2t)+2K(t)$, and $K_{-}=K(2t)-2K(t)$.
\begin{figure}
\includegraphics[width=3in]{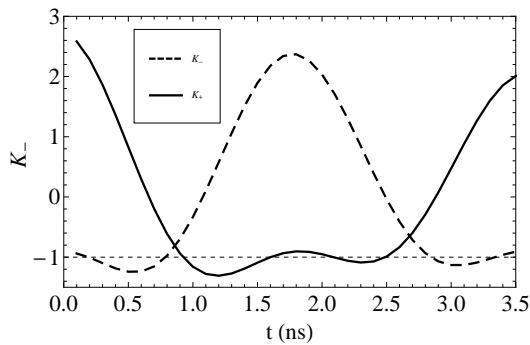}
\caption{Two types of LG inequalities. The fine-structure splitting
$S=3$ $\mu$eV, background noise g=0, the gate width $\omega=50$ ps
and the temperature $T=5$ K.}
\end{figure}

Our result is shown in Fig. 2. We can see that the curve of the LG
oscillates with time, what's more, the amplitude of the curve decays
when the time increases. By comparing the two types of LG
inequalities, we find that the $K_{-}$ reaches the classical limit
$-1$ first. As we know, when the time delay of the photon pairs
increase, the biphoton coincidence decreases. Therefore, $K_{-}$ is
easier to be tested in the experiment and we discuss $K_{-}$ for the
following results.
\begin{figure}
\includegraphics[width=3in]{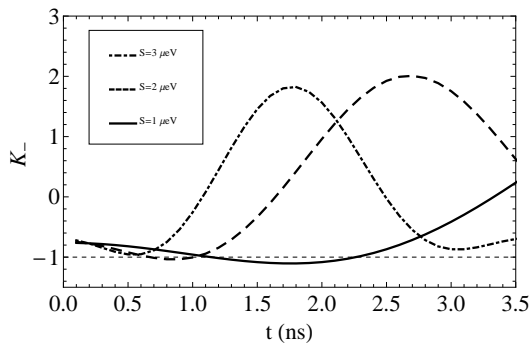}
\caption{The relationship between fine-structure splitting and
Leggett-Garg inequality. In the conditions of the background noise
$g=0.3$, the gate width $\omega=50$ ps and the temperature $T=5$ K.
The violation of the LG inequality becomes easier when the
fine-structure splitting decreases.}
\end{figure}
\begin{figure}
\includegraphics[width=3in]{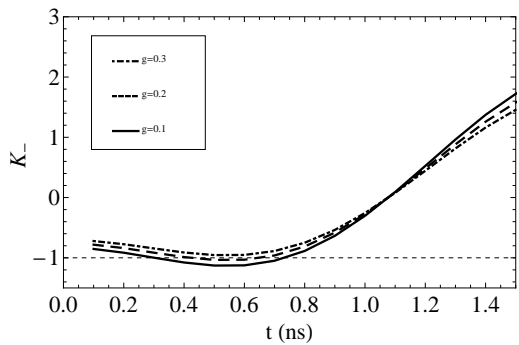}
\caption{The relationship between background noise and the LG
inequality. The fine-structure splitting $S=3$ $\mu$eV , the gate
width $\omega=50$ ps and the temperature $T=5$ K.}
\end{figure}
\begin{figure}
\includegraphics[width=3in]{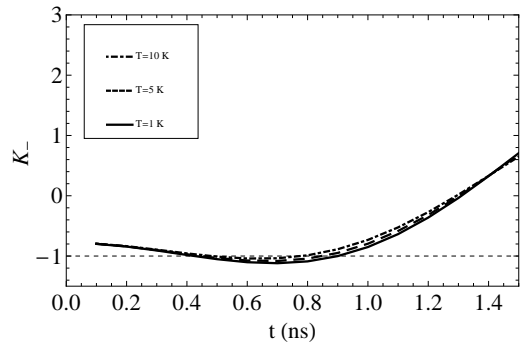}
\caption{The relationship between temperature and the LG inequality.
The fine-structure splitting $S=2.5$ $\mu$eV, the background noise
$g=0.2$ and the gate width $\omega=50$ ps.}
\end{figure}

Then we analysis the relationship between the LG inequality and the
fine-structure splitting. The result is presented in Fig. 3. In this
case, the background noise $g=0.3$ , the gate width $\omega=50$ ps
and the temperature $T=5$ K. When the fine-structure splitting is
small, the LG inequality is violated easily. However, with the
increase of the fine-structure splitting, the violation of LG
inequality becomes increasingly weak. It can be seen that $K_{-}$ do
not violate the classical limit $-1$ when the fine-structure
splitting becomes large enough. This implies that when the
fine-structure splitting becomes large, the evolution process can be
described by classical realistic theory.

Further more, the influence of background noise and temperature are
investigated. As shown in Fig. 4 and Fig. 5, when the background
noise and temperature decrease, the curve of the LG inequality goes
below the classical limit $-1$ easily. The transition from quantum
process to classical process is also been seen clearly.

At last, it should be noticed that there is an approximation in our
model. We use the state of photon emitted to determine the state of
quantum dot, this is true when the fine structure splitting is zero.
In real situation, we need the fine structure splitting to be
nonzero in order to introduce the evolution, the partial
distinguishability of spectrums between $|H\rangle$ and $|V\rangle$
will introduce errors. However, it can be easily verified that the
errors only make the LG inequality more difficult to be violated,
i.e., the requirement is not loosen in our model.

In summary, we have investigated the violation of the LG
inequalities in quantum dot system. When the fine-structure
splitting of the quantum dot system, background noise and the
temperature become small, we achieve the maximal violation. Better
results may be obtained, when we change the method by which we get
the initial state. Finally, we can see clearly from the results that
the LG inequalities can be used as a criterion to identify the
boundary of the classical realistic description.

This work was supported by National Fundamental Research Program,
National Natural Science Foundation of China (Grant Nos. 60921091
and 10874162).
\bibliographystyle{}

\end{document}